\documentclass[aps,preprint]{revtex4}
\usepackage{amsfonts}
\usepackage{amsmath}
\usepackage{amssymb}
\usepackage{graphicx}
\usepackage{subfigure}
\usepackage{mathrsfs, mathtools}
\usepackage{pdfpages}
\usepackage{float}
\usepackage{color}
\usepackage{xcolor}
\usepackage{caption}
\usepackage{lineno,hyperref}
\hypersetup{colorlinks =true, allcolors = blue}
\providecommand{\U}[1]{\protect\rule{.1in}{.1in}}

\begin{document}
\preprint{HEP/123-qed}
\title{Accretion, greybody factor, quasinormal modes, power spectrum, sparsity of Hawking radiation, and weak gravitational lensing of a minimum measurable length inspired Schwarzchild black hole}
\author{Himangshu Barman}
\affiliation{Department of Physics, Hooghly Mohsin College, Chinsurah, Hooghly - 712101, West Bengal, India}
\author{Ahmad Al-Badawi}
\email{ahmadbadawi@ahu.edu.jo}
\affiliation{Department of Physics, Al-Hussein Bin Talal University, P. O. Box: 20, 71111,
Ma'an, Jordan.}
\author{Sohan Kumar Jha}
\email{sohan00slg@gmail.com(corresponding author)}
\affiliation{Department of Physics, Chandernagore College, Chandernagore, Hooghly, West Bengal, India}
\author{Anisur Rahaman}
\email{manisurn@gmail.com} \affiliation{Department of Physics,
Durgapur Government College, Durgapur, Burdwan 713214, West
Bengal, India.}

\begin{abstract}
In this manuscript, we delve into an analytic and numerical probe of shadow with different accretion models, quasinormal modes, Hawking radiation, and gravitational lensing to study observational impacts of quantum effect introduced throughh linear-quadratic GUP(LQG). Our investigation reveals that the shadows of LQG modified black holes are smaller and brighter than Schwarzschild black holes.  To examine the impact of the quantum correction on the quasinormal
mode, linear-quadratic GUP modified black holes are explored under scalar and
electromagnetic field perturbation. Here, linear-quadratic GUP is
used to capture quantum corrections. It is observed that the
incorporation of quantum correction by linear-quadratic GUP alters
the singularity structure of the black hole. To compute the
quasinormal modes of this linear-quadratic GUP-inspired
quantum-corrected black holes, we compute the effective potential
generated under the perturbation of scalar and electromagnetic
field, and then we use the sixth-order WKB approach in conjunction
with the appropriate numerical analysis. We find that the greybody factor decreases with the GUP parameter $\alpha$ implying that the probability of transmission decreases with the GUP parameter. The total power emitted by LQG modified black hole is found to be greater than that emitted by Schwarzschild black hole. Finally, we study weak gravitational lensing and make a comparison with quadratic GUP and linear GUP modified black holes.
\end{abstract}
\volumeyear{ }
\eid{ }
\date{\today}
\received{}

\maketitle
\tableofcontents
\section{Introduction}
A black hole is a celestial body predicted through Einstein's general theory of relativity. A crucial feature of black holes is
Hawking radiation \cite{HAWKING1, HAWKING2} which although has not
been detected yet, has always attracted a lot of attention
from the scientific community. After the direct detection of
gravitational waves from black hole mergers, along with the
discovery of the first image of the supermassive black hole at the
center of the supergiant elliptical galaxy M87 through the Event
Horizon Telescope (EHT) project \cite{EHT1, EHT2, EHT3, EHT4,
EHT5, EHT6}, several significant investigations on various aspects
of black holes have gained more attention. The characteristics of
an isolated black hole includes its mass, angular momentum, and
charge. However, the idea of an isolated black hole is an idealized
concept. The cosmos does not contain isolated black
holes. These black holes are expected to interact continually with
both their external environment and their inside complicated
matter fields.

The theory of general relativity asserts that the perturbation of
space-time associated with black holes is expected to accompany the
emergence of quasinormal modes (QNMs) in an essential way. As long
as black holes undergo perturbation, the general theory of
relativity, as well as alternative theories of gravity, suggest that
they face the influence of some effective potential around them
and it is instructive to look for them. A natural way to respond
to the perturbation is indeed the emission of gravitational waves.
The process of evolution of gravitational waves can be divided
into three stages \cite{KONOR}: a relatively short period of an initial outburst of radiation that depends strongly on the initial
conditions, followed by a longer period of exponentially decaying
ringdown phase at the intermediate time dominated by quasinormal
mode, which depends entirely on the parameters of the black hole,
QNMs and finally a suppression over a longer period of time
by a power law or exponential late-time tails. QNMs are
characteristic frequencies with a non-vanishing imaginary part,
which encodes the information on how black holes relax after being
exposed to perturbation. The quasinormal frequencies depend on the
details of the geometry and the type of the perturbation
(scalar, vector (electromagnetic), tensor (gravitational), or
fermionic) it encounters, the initial conditions have very little
to do with it. This implies that black holes are interacting
continually with the external source field; hence, it is
anticipated that real black holes will be in a perturbed
condition.

Black hole perturbation theory \cite{PARTERBATION1,
PARTERBATION2,PARTERBATION3,PARTERBATION4,PARTERBATION5,PARTERBATION6}
and the determination of QNMs has become pertinent to have a
better understanding of black holes. Physicists are typically more
concerned about the ringdown stage among these
three, and consequently, the research community has been paying a
great deal of attention to QNMs. First and foremost, there is
curiosity from an experimental perspective since there is a chance
that QNMs could one day be detected by gravitational antennas like
LIGO, VIRGO, and LISA \cite{LIGO1, LIGO2, LIGO3, LIGO4, VIRGO}.
Such detections would provide hints to determine the physical
characteristics of the black holes since the QNMs solely depend on
the features of the black holes. On the other hand, numerous
studies have concentrated on analyzing QNMs of black holes with a
negative cosmological constant due to the well-known relationship
between AdS/CFT duality \cite{13, 14}. The hypothesis made by Hod
\cite{15} linking the quantum characteristics of Schwarzschild
black holes and asymptotic QNMs was another factor igniting
interest in QNMs. In that vein, there are numerous publications
targeted at computing asymptotic QNM frequencies.
Konoplya and Zhidenko provide an outstanding analysis of QNMs \cite{16}. The
recent research connecting QNM to several facets of black hole
physics is exciting. For instance, some recent research \cite{17,
18} investigated the relationship between QNMs and veiled
conformal symmetry. Stefanov et al. \cite{19} published another
intriguing research on the relationship between gravitational
lensing and QNMs. Even naked singularities, which are not black
holes, have been researched from a QNM perspective \cite{20}.
There have been a lot of interesting studies on black hole QNMs in
the literature [\citenum{KONO}-\citenum{36}]. Even though there are
many indirect ways to identify black holes in the universe,
perturbed black holes' gravitational waves will create distinctive
"fingerprints" that enable scientists to confirm their presence
directly. In particular, Cardoso suggests using gravitational wave
echoes as a brand-new characteristic of unusual compact objects
\cite{37}. Gravitational wave echoes were later found in the late
stage of quasinormal ringing when QNM was explored in various
space-time contexts \cite{38, 39, 40, 41, 42, 43, 44, 45, 47, 48,
49, 50}. They are eager to use it to distinguish between various
compact objects.

In theoretical physics, the generalized uncertainty
principle (GUP) presents a remarkable idea known as the minimum
measurable length, which has a strong connection to string theory
\cite{KEMPF, DAMIT, MAGG1, MAGG2}. Consequently, it offers a
technique to incorporate the quantum gravity effect. The concept
of minimum measurable length has considerable application to
the physics of black hole \cite{BHA1, BHA2, BHA3, BHA4, BHA5,
BHA6, BHA7}. These objects in the universe are important
for classical and quantum gravity. However, the theory of
quantum gravity has yet to be available in a mature form. Therefore, a fruitful
area of research is the study of the characteristics of
black holes that capture quantum gravity correction. A
promising method of capturing quantum correction results from this
novel idea that results from the GUP approach. The Schwarzschild
radius gets modified when quantum correction inspired by GUP is
employed \cite{SCRMOD}, and that eliminates the singularity of the
black hole metric. Therefore, the study of QNM with GUP-inspired
quantum-corrected Schwarzschild black hole would be instructive
and of interest.

Quadratic-GUP (QG) \cite{KEMPF},
linear-quadratic GUP (LQG) \cite{LQGUP}, extended GUP
\cite{EXTGUP1, EXTGUP2}, exponential GUP \cite{EXPGUP} etc., are
generally used. Although these GUPs are of different origins, so
far as constructions of these are concerned, the main service
these GUPs render is common: providing quantum gravity
correction to the physical system. A few recent works with LQG and QG are
\cite{LQCOSMO, LQT, GWA, TFM, QGUP}. Another significant astrophysical phenomenon related to emission from a black hole is 
Hawking radiation. Hawking, incorporating quantum consequences, proved that black holes radiate \cite{HAWKING}. A pair production occuring near event horizon produces two particles-one is consumed by the black hole and another one moves away from it constituting the Hawking radiation [\citenum{HH}-\citenum{HC}]. The hawking temperature can be obtained using different techniques [\citenum{SW}-\citenum{SI}]. An important quantity associated with the Hawking radiation is the greybody factor. The greybody factor underlines the transmission probability of the Hawking radiation. Various techniques for calculating greybody bounds and their applications can be found in [\citenum{qn32}-\citenum{WJ}].\\
Accretion plays a key role in explaining various astrophysical phenomena, such as
electromagnetic radiation in the form of gamma-ray bursts, tidal disruption events, and
X-ray binaries. In this article, we intend to find the signature of the quantum correction
on the shadow with static and infalling accretion. Since the shadow images bear the imprints of underlying
space-time and the accretion process, this will provide us with a unique opportunity to gauge the impact of
quantum correction on such a significant astrophysical phenomenon. Some notable works in this regard are [\citenum{RN}-\citenum{XX}]. The gravitational lensing is one of the important predictions of GR. It has been subject of intense study for quite sometime now for its importance in determining mass of galaxies and clusters and discovering existence of dark matter and dark energy [\citenum{251}-\citenum{271}]. There are two types of GL: sttrong and weak. Gibbons and Werner, with the help of the Gauss-Bonnet theorem and the optical geometry of black hole \cite{CARMO}, provided a technique for obtaining the deflection angle in weak GL \cite{GW}. This technique was extended to stationary spacetimes by Werner in his work \cite{WERNER}. Ishihara et. al. extended the approach for finite distances \cite{ISHIHARA1, ISHIHARA2} which was later applied in their work by Ono et. al. \cite{ONO1,ONO2, ONO3}. Higher order terms was obtained in \cite{CRISNEJO}. Some of the notable studies in weak GL for different spacetimes are [\citenum{411}-\citenum{511}].\\
The rest of the manuscript is organized as follows. In Sec. (II), we introduce LQG modified metric and study shadow images and intensities with static and infalling accretion. Sec. (III) is devoted to the study of quasinormal modes, and in Sec. (IV), time profiles of scalar and electromagnetic perturbations are provided. Sec. (V) deals with the greybody factor, and Sec. (VI) is where power spectrum and sparsity are studied. Weak gravitational lensing is studied in Sec. (VII). We end our article with conclusions and discussion in Sec. (VIII).
\section{General description of Heisenberg algebra with the
linear and quadratic term in momentum} In mathematical form, the
celebrated Heisenberg algebra is expressed by the expression
\begin{equation}
 [x, p]= i\hbar\label{HCOM}.
\end{equation}
The above algebra is  associated with the fundamental uncertainty
relation
\begin{equation}
\Delta x \Delta p \ge \frac{\hbar}{2}. \label{QCOM}
\end{equation}
There are several generalizations of the Heisenberg algebra. The
generalization with quadratic terms in momentum reads
\begin{equation}
[x, p]= i\hbar(1+ \tilde{\alpha}^2 p^2). \label{COMC}
\end{equation}
The generalization of Heisenberg algebra with a combination of
linear and quadratic terms in momentum  came in the form
\begin{equation}
[x, p]= i\hbar(1-2\tilde{\alpha} p + 4\tilde{\alpha}^2 p^2).
\label{COM}
\end{equation}
 The uncertainty relation associated with  the generalized
algebra (\ref{COM}) is given by
\begin{eqnarray}
\Delta x \Delta p &\ge&\hbar(1-2\tilde{\alpha} <p> + 4\tilde{\alpha}^2 <p^2>) \nonumber \\
&\ge& \frac{\hbar}{2}[(1+
\frac{\tilde{\alpha}}{\sqrt{<p^2>}}+4\tilde{\alpha}^2)\Delta p^2
+4\tilde{\alpha}^2<p>^2-2\tilde{\alpha}\sqrt{<p>^2}].
\end{eqnarray}
Here $\tilde{\alpha} = \frac{\alpha}{M_{pl}
c}=\alpha\frac{L_{pl}}{\hbar}$, where $M_{pl} c^2 = 10^{19}GeV$,
 the Planck length $L_{pl} = 10^{-35}m$, and $c$ is the velocity of light.
Unlike the quadratic generalization, the concept of maximum
momentum along with the minimum length is admissible here and
here lies the fundamental difference between these two. The
minimum length and maximum momentum admissible to this deformed
algebra respectively are
\begin{equation} \delta x\ge\delta x_{min} \approx \alpha
l_p,~~~ \delta p\le\delta p_{max} \approx \frac{M_{pl} c}{\alpha}.
\end{equation}
These are the necessary inputs for this generalized uncertainty
relation having a combination of linear and quadratic terms of
momentum to deal with it. We, therefore, are in a position to
apply it to the model we have considered for extension. To be
precise we will apply this LQG to study the quantum-corrected QNMs
with an emphasis on how it deviated when quantum correction was
captured from QG. For the LQG, the uncertainty in position is given
by
\begin{eqnarray}
\Delta x> \frac{\hbar}{\Delta
p}-2\frac{\alpha\hbar}{M_{pl}c}+4\hbar(\frac{\alpha}{M_{pl}c})^2\Delta
p, \label{equ01}
\end{eqnarray}
where $M_{pl}$ is Planck mass. If we consider $\Delta x\rightarrow
r$, $\Delta p\rightarrow Mc$, then we get
\begin{eqnarray}
r> {r}_{C}\equiv
\frac{\hbar}{Mc}[1-2\alpha\frac{M}{M_{pl}}+4\alpha^2(\frac{M}{M_{pl}})^2],
\end{eqnarray}
where $r$ stands for radial separation and $M$ represents the mass
of the body under consideration. This expression $\tilde{r}_{C}$
might be regarded as a modified Compton wavelength, the last term
of which is representing a quantum  correction \cite{SCRMOD}
Consequently, the black hole horizon in this situation also gets
modified. We therefore have
\begin{eqnarray}
\tilde{r}> \tilde{r}_{s}= \frac{4\alpha^2
M\hbar}{M_{pl}^{2}c}\left [
1-\frac{1}{2\alpha}(\frac{M_{pl}}{M})+\frac{1}{4\alpha^2}(\frac{M_{pl}}{M})^2\right
]. \label{equ02}
\end{eqnarray}
Since the free constant of Eq.(\ref{equ01}) is connected to the
the first term, the above two equations can be written down as
\begin{eqnarray}
\tilde{r}_C=\frac{\alpha \hbar}{Mc} \left
[1-\sqrt{\frac{2}{\alpha}}(\frac{M}{M_{pl}})+\frac{2}{\alpha}(\frac{M}{M_{pl}})^2\right],
\end{eqnarray}
and
\begin{eqnarray}
\tilde{r}_{s}= \frac{2M\hbar}{M_{pl}^{2}c}\left
[1-\sqrt{\frac{\alpha}{2}}(\frac{M_{pl}}{M})+\frac{\alpha}{2}(\frac{M_{pl}}{M})^2\right
],
\end{eqnarray}
where $\tilde{r}_{s}$ the modified Schwarzschild radius, and
$\alpha$ is a positive dimensionless parameter. For numerical
computation, we have taken $\hbar=c=1$. Note that when modification
due to QG is taken into account
\begin{equation}
r_{s}^{'}=\frac{2GM}{c^2}\left(1+\frac{\alpha}{2}\frac{M^{2}_{pl}}{M^2}\right).
\end{equation}
Therefore, the QG-corrected Schwarzschild metric will be given by
\begin{eqnarray}
\text{d}s^{2}=B(r)\text{d}t^{2}-\frac{1}{B(r)} \text{d}
r^{2}-r^{2}d\Omega^2, \label{equ1}
\end{eqnarray}
but the LQG corrected Schwarzschild metric is

\begin{eqnarray}
\text{d}s^{2}=A(r)\text{d}t^{2}-\frac{1}{A(r)} \text{d}
r^{2}-r^{2}d\Omega^2 , \label{equ1}
\end{eqnarray}
where
\begin{eqnarray}
d\Omega^2=(\text{d}\theta ^{2}+\sin^{2}\theta \text{d}\varphi
^{2})~~~ A(r)=1-\frac{\tilde{r}_s}{r} ~~~B(r)=1-\frac{r'_s}{r}.
\label{equ2}
\end{eqnarray}
There is a fair amount of literature on the application of
different types of GUP  in different branches of physics
\cite{EXTGUP1, EXTGUP2, EXPGUP, TFM, AFALI1, AFALI2, AFALI3,
AFALI4, SUN1, SUN2, SUN3, SUN4}. To study null geodesics, we confine ourselves to the equatorial plane without loss of generality, as the metric under consideration is spherically symmetric. The Lagrangian corresponding to the metric (\ref{equ2}) is
\begin{equation}
\mathscr{L}=A(r)\dot{t}^2-\frac{\dot{r}^2}{A(r)}-r^{2}\dot{\varphi}^2.
\end{equation}
With the help of the definition $p_q=\frac{\partial \mathscr{L}}{\partial \dot{q}}$, where $p_q$ is the conjugate momentum to the coordinate $q$, we obtain
\begin{eqnarray}\nonumber
p_t&=&\frac{\partial  \mathscr{L}}{\partial \dot{t}}=A(r)\dot{t}=\mathcal{E}, \\\nonumber
p_r&=&\frac{\partial  \mathscr{L}}{\partial \dot{r}}=-\frac{\dot{r}}{A(r)}, \\
p_\varphi&=&\frac{\partial  \mathscr{L}}{\partial \dot{\varphi}}=-r^2\dot{\varphi}=-\mathcal{L}.
\end{eqnarray}
Here, the dot corresponds to differentiation with respect to an affine parameter $\lambda$. The above equations yield the differential equation of motion for the null geodesics as
\begin{eqnarray}\nonumber
\left(\frac{d r}{d \lambda}\right)^{2} \equiv \dot{r}^{2}&=&\mathcal{E}^{2}-\frac{\mathcal{L}^{2} A(r)}{r^2}\\
&=&\mathcal{E}^{2}-V(r),
\end{eqnarray}
We employ the following conditions to obtain the radius of an unstable photon orbit:
\begin{equation}
V(r_p)=0,\quad\frac{d V}{d r}|_{r=r_{p}}=0,\quad\text{and}\quad \frac{\partial^2V}{\partial r^2}|_{r=r_{p}}<0.
\label{condition}
\end{equation}
These conditions yield the photon radius given by
\begin{equation}
r_{p}=\frac{3}{2} \frac{2 M}{M_{\text{pl}}^2}\left(\frac{\alpha  M_{\text{pl}}^2}{2 M^2}-\frac{\sqrt{\alpha } M_{\text{pl}}}{\sqrt{2} M}+1\right).
\label{rp}
\end{equation}
The impact parameter corresponding to the photon radius, which is also the shadow radius of an asymptotic observer, is given by
\begin{eqnarray}\nonumber
b_p=R_{s}&=&\frac{\mathcal{L}}{\mathcal{E}}=\frac{r_p}{\sqrt{A(r_p)}}\\
&=&\frac{9}{2} \frac{2 M}{ M_{\text{pl}}^2}\left(\frac{\alpha  M_{\text{pl}}^2}{2 M^2}-\frac{\sqrt{\alpha } M_{\text{pl}}}{\sqrt{2} M}+1\right).
\label{bp}
\end{eqnarray}
In Eqs. (\ref{rp}, \ref{bp}), we have put $\hbar=c=1$. Eqs. (\ref{rp}, \ref{bp}) clearly indicate the impact of the GUP parameter $\alpha$ on the motion of photons in the background of the GUP-modified black hole. To have a better understanding of the impact, we tabulate numerical values of event horizon $r_h$, photon radius, and corresponding impact parameter. Table (\ref{shadowradius}) clearly shows that both the radii and the impact parameter decrease as we increase the GUP parameter $\alpha$. 
\begin{table}[!htp]
\centering
\caption{Various values of photon radius and shadow radius for various values of $\alpha$.}
\begin{tabular}{|c|c|c|c|c|c|c|c|c|c|c|c|}
\hline
 $\alpha$  & 0. & 0.2 & 0.4 & 0.6 & 0.8 & 1. & 1.2 & 1.4 & 1.6 & 1.8 & 2. \\
\hline
$ r_h $ & 8. & 6.93509 & 6.61115 & 6.40911 & 6.27018 & 6.17157 & 6.10161 & 6.05336 & 6.02229 & 6.00527 & 6. \\
\hline
$ r_p$ & 12. & 10.4026 & 9.91672 & 9.61366 & 9.40527 & 9.25736 & 9.15242 & 9.08004 & 9.03344 & 9.0079 & 9. \\
\hline
$ b_p$ & 20.7846 & 18.0179 & 17.1763 & 16.6514 & 16.2904 & 16.0342 & 15.8525 & 15.7271 & 15.6464 & 15.6021 & 15.5885 \\
\hline
\end{tabular}
\label{shadowradius}
\end{table}
Black hole shadows provide a viable window to probe further the impact of the GUP parameter on astrophysical observables. To this end, we will be investigating the optical appearances of the black hole with static as well as infalling accretion flows. The prime focus of our investigation is to observe the impact of $\alpha$. 
\subsection{Black hole shadow and photon sphere with the static spherical accretion}
We consider static optically and geometrically thin accretion flow. The intensity measured by a distant observer at the frequency $\nu_{obs}$ along any ray path is given by \cite{mj97, bambi13}
\begin{equation}
\label{intensity}
I(\nu_{obs})=\int_{ray}g^{3}j(\nu_{em})dl_{prop},
\end{equation}
where $g=A(r)^{1/2}$ is the redshift factor, $\nu_{em}$ is frequency of photon emitted, and $\nu_{obs}$ is the observed photon frequency. , and is proper length differential as measured by an observer in the frame moving with the matter. By assuming monochromatic nature of emission with rest-frame frequency $\nu_{t}$, the emissivity per unit volume, $j(\nu_{em})$ and the proper length differential, $dl_{prop}$, are given by
\begin{eqnarray}
j(\nu_{em})\propto \frac{\delta(\nu_{em}-\nu_{t})}{r^2},~~dl_{prop}=\sqrt{A(r)^{-1}+r^{2}\Big(\frac{d\varphi}{dr}\Big)^{2}}dr.
\end{eqnarray}
With the help of the above equations, we can write
\begin{equation}
\label{3-1-3}
I(\upsilon_{obs})=\int_{ray}\frac{A(r)^{3/2}}{r^{2}}\sqrt{A(r)^{-1}+r^{2}\Big(\frac{d\varphi}{dr}\Big)^{2}}dr.
\end{equation}
Photons with impact parameter $b_p$ revolve around a black hole several times before reaching a distant observer. Photons with impact parameter $b>b_{p}$ get deflected towards the distant observers, and those with $b<b_{p}$ get swallowed by the black hole. We plot the observed intensity with respect to the impact parameter b for different values of $\alpha$. It shows that the intensity increases with $b$ reaching its maximum at $b_{p}$ and then falls off rapidly. The peak of intensity occurs at $b_p$ because the photon revolves around the black hole several times. The peak value of intensity increases with $\alpha$, but the position of the peak value shifts towards the left as we increase the GUP parameter. This is consistent with Table (\ref{shadowradius}). Shadows of the black hole for different values of $\alpha$ are shown in Fig. (1). The bright ring with the strongest luminosity in each shadow is the photon ring. The inner region of the shadow is not entirely black because a tiny fraction of photons can escape a black hole. It is also to be noted that the luminosity of the photon ring gets enhanced due to the GUP parameter, whereas the shadow size decreases. 
\begin{center}
\includegraphics[width=6cm,height=6cm]{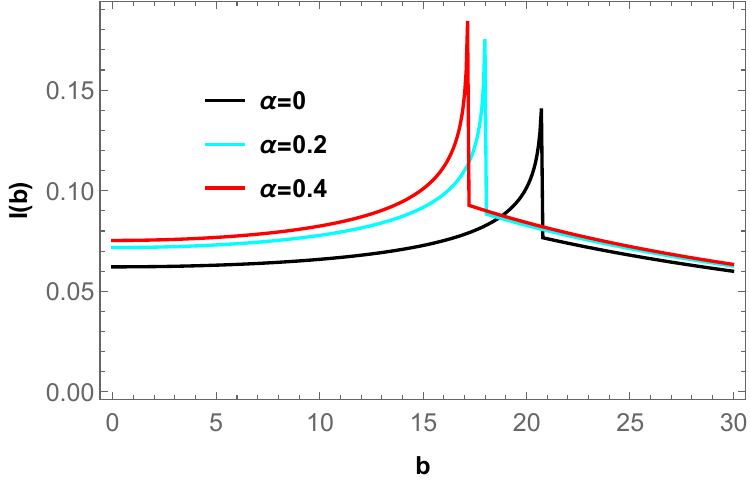}
\includegraphics[width=6cm,height=6cm]{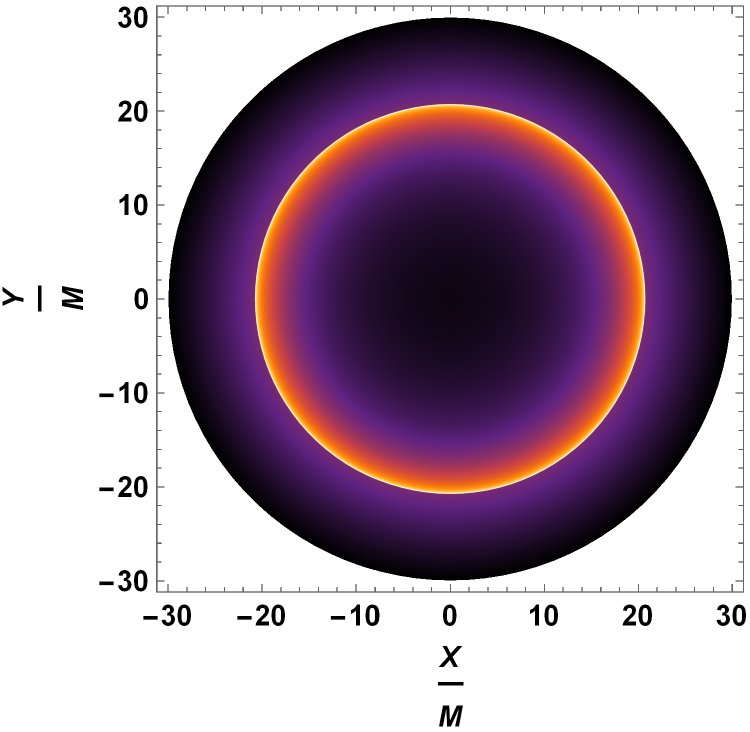}
\includegraphics[width=0.5cm,height=6cm]{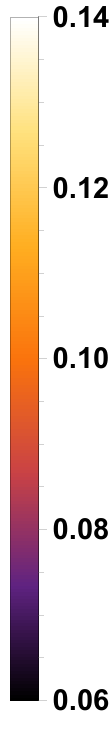}
\includegraphics[width=6cm,height=6cm]{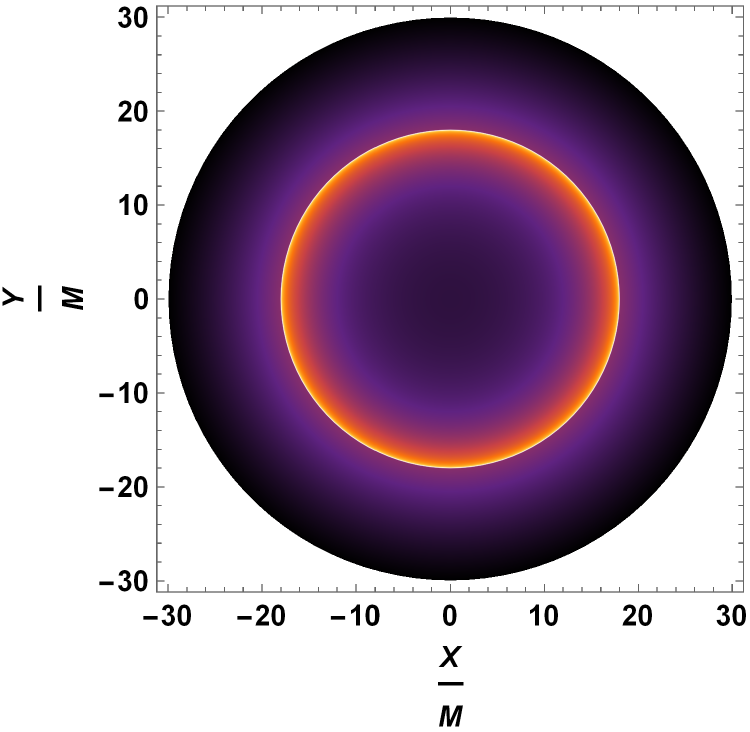}
\includegraphics[width=0.5cm,height=6cm]{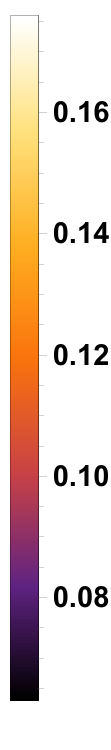}
\includegraphics[width=6cm,height=6cm]{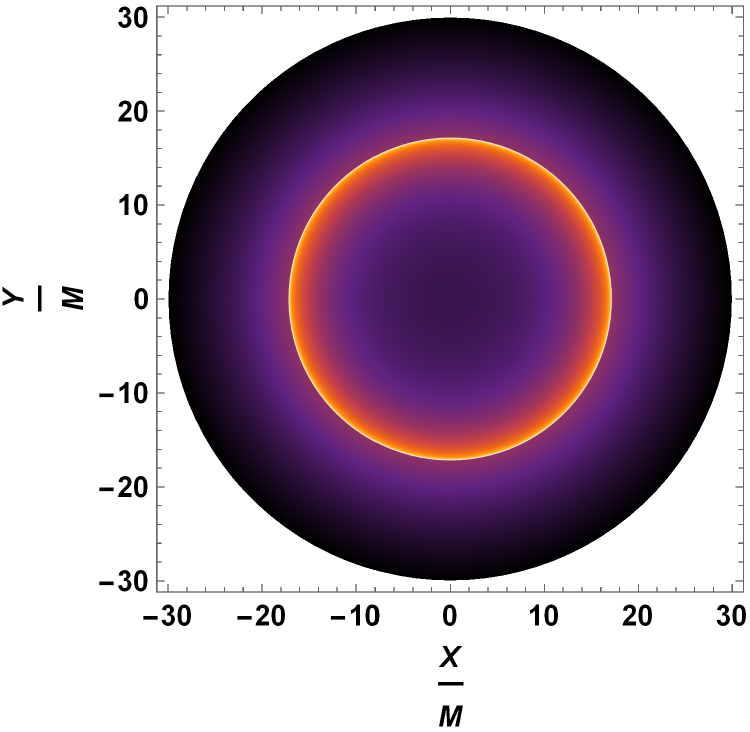}
\includegraphics[width=0.5cm,height=6cm]{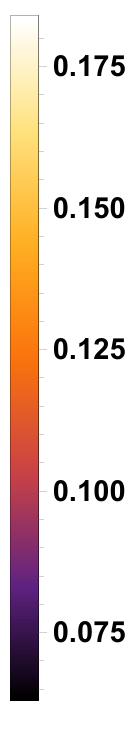}
\parbox[c]{15.0cm}{\footnotesize{\bf Fig~1.}  
Observed intensity and black hole shadows for static spherical accretion. The upper right is for $\alpha=0$, the lower left one is for $\alpha=0.2$, and the lower right one is for $\alpha=0.4$.}
\label{staticshadow}
\end{center}
\subsection{Black hole shadow and photon sphere with the infalling spherical accretion}
Here, we consider an infalling spherical accretion since most accretions are moving in the real world. The observed intensity is also given by equation (\ref{intensity}) with the redshift factor
\begin{equation}
g=\frac{k_{\gamma}u_{obs}^{\gamma}}{k_{\zeta}u_{em}^{\zeta}}.
\end{equation}
In the above equation, $k^{\mu}\equiv \dot{x}_{\mu}$ is photon four-velocity, $u_{obs}^{\mu}\equiv(1,0,0,0)$ is distant observer four-velocity, and $u_{em}^{\mu}$ is accretion four-velocity. With $k_{t}=1/b$ combined with $k_{\mu}k^{\mu}=0$ we obtain
\begin{equation}
\frac{k_{r}}{k_{t}}=\pm\frac{1}{A(r)}\sqrt{1-\frac{b^{2}A(r)}{r^{2}}},
\end{equation}
where the sign + is when photons approach the black hole, and the - sign is when photons move away from the black hole. The accretion four-velocity is
\begin{eqnarray}
u_{em}^{t}=\frac{1}{A(r)},~~~~~u_{em}^{\theta}=u_{em}^{\varphi}=0,~~~~~u_{em}^{r}=-\sqrt{1-A(r)}.
\end{eqnarray}
The redshift factor, with the help of the above equations, becomes
\begin{equation}
g_{f}=\frac{1}{u_{em}^{t}+k_{r}/k_{em}u_{em}^{r}}.
\end{equation}
We can write the proper distance as
\begin{equation}
dl_{prop}=k_{\mu}u_{em}^{\mu}d\zeta=\frac{k_{t}}{g_{f}|k_{r}|}dr.
\end{equation}
Thus, for the infalling spherical accretion, we have
\begin{equation}
\label{3-2-6}
I(\nu_{obs})\propto \int_{ray}\frac{g_{f}^{3}k_{t}dr}{r^{2}|k_{r}|}.
\end{equation}
We intend to investigate the impact of the GUP parameter on the observed luminosity and optical appearance of black holes with infalling accretion. One can observe that the peak value of intensity still occurs at $b=b_p$, but the variation of intensity with respect to the impact parameter is smooth compared to the static case. It can also be observed that the peak value of intensity is smaller than that of the static case, rendering the black hole darker with infalling accretion. This observation is evident from the black shadows provided in Fig. (2). The central region is much darker, and the brightness of the photon ring is far less when compared to the static case. The Doppler effect is the reason behind this difference. It can also be observed that the shadow radius and photon sphere do not change with accretion flow. This shows that the shadow radius and photon sphere depend only on the underlying space-time.
\begin{center}
\includegraphics[width=6cm,height=6cm]{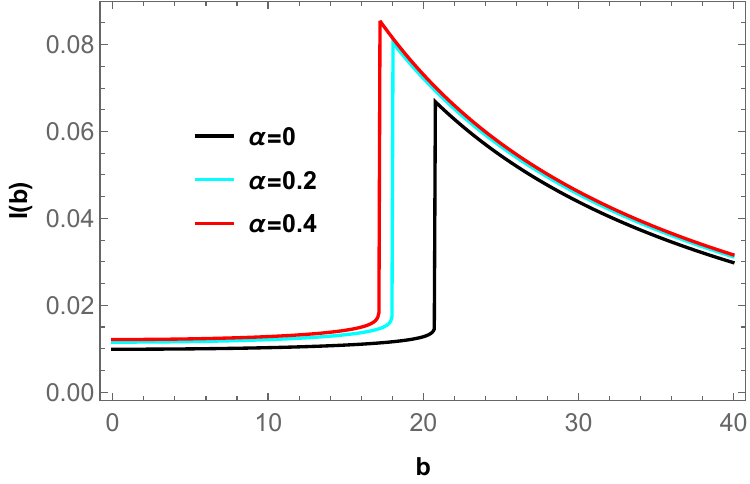}
\includegraphics[width=6cm,height=6cm]{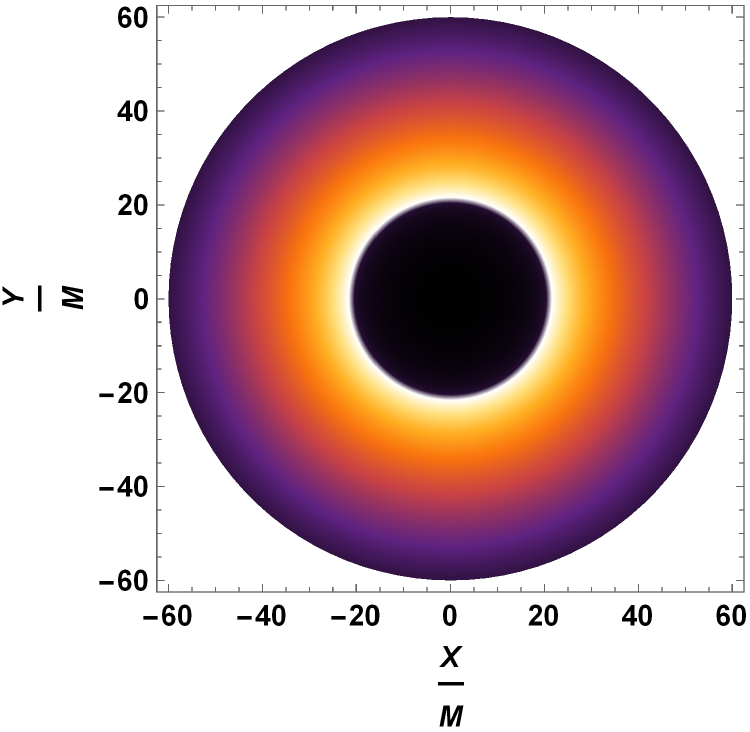}
\includegraphics[width=0.5cm,height=6cm]{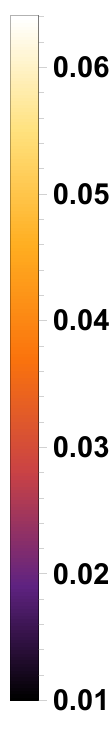}
\includegraphics[width=6cm,height=6cm]{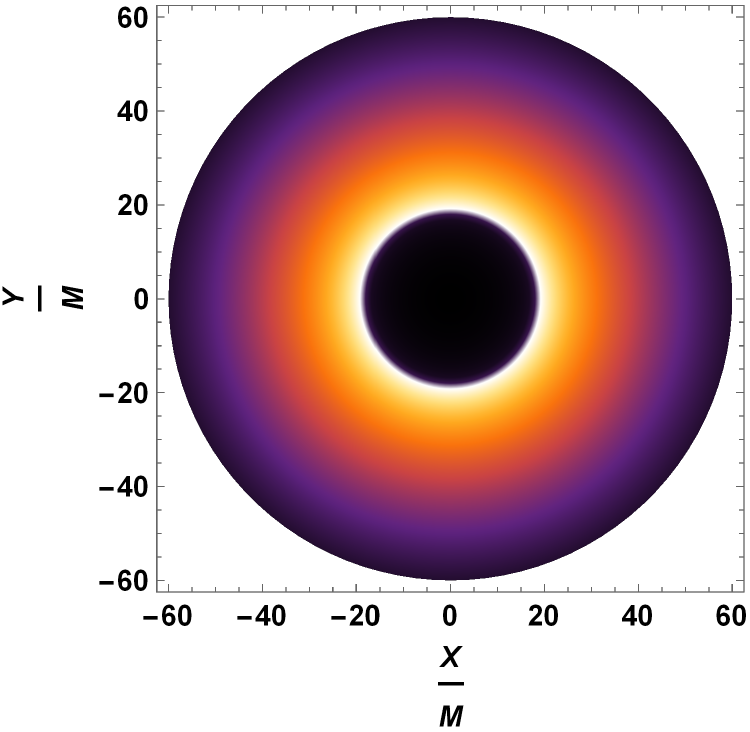}
\includegraphics[width=0.5cm,height=6cm]{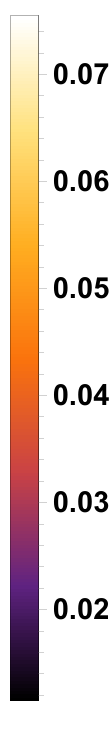}
\includegraphics[width=6cm,height=6cm]{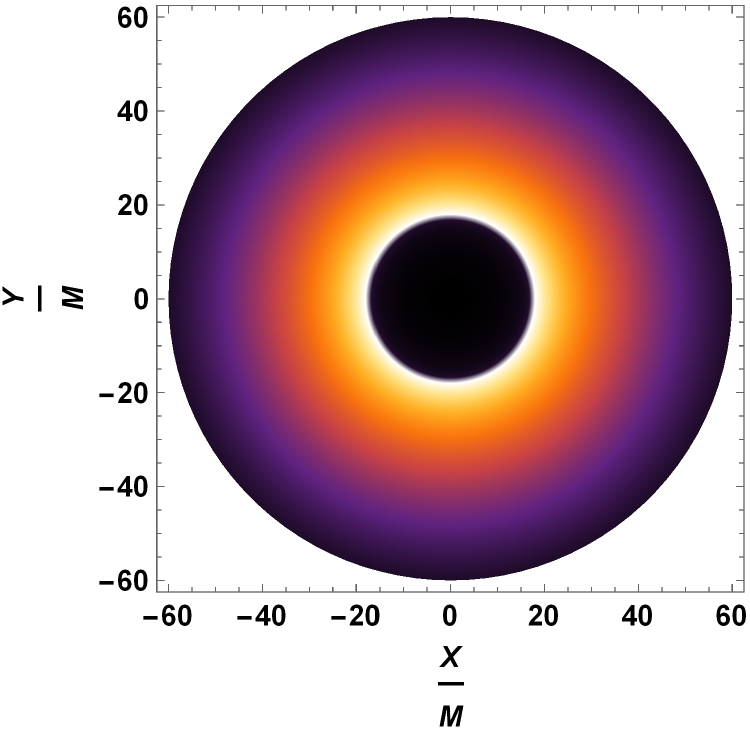}
\includegraphics[width=0.5cm,height=6cm]{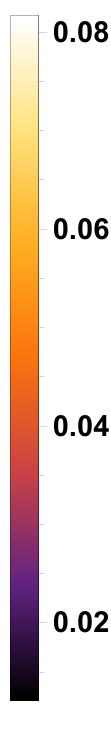}
\parbox[c]{15.0cm}{\footnotesize{\bf Fig~2.}  
Observed intensity and black hole shadows for infalling spherical accretion. The upper right is for $\alpha=0$, the lower left one is for $\alpha=0.2$, and the lower right one is for $\alpha=0.4$.}
\label{infallingshadow}
\end{center}

\section{Quasinormal Frequencies}

\begin{figure}
    \centering
{{\includegraphics[width=8.5cm]{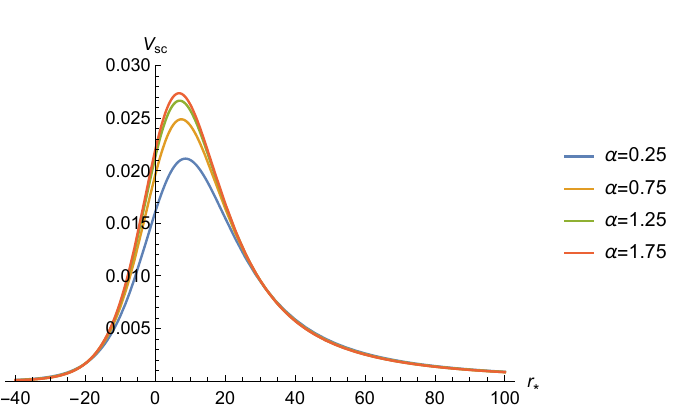} }}\qquad
    {{\includegraphics[width=7cm]{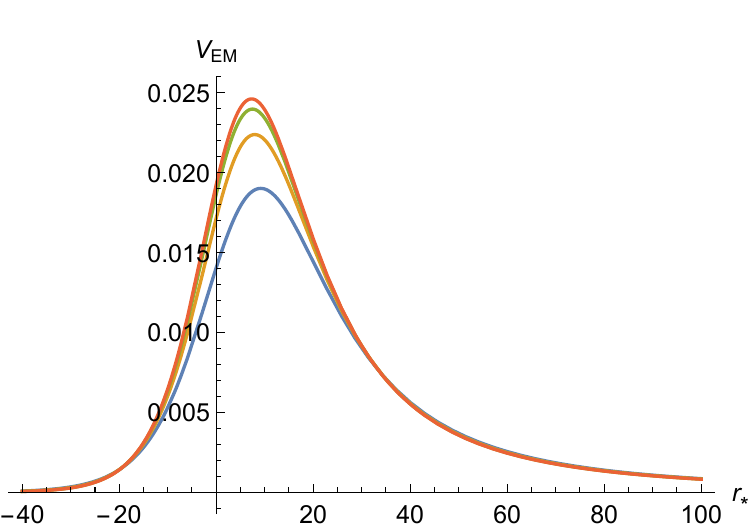}}}
    \caption{ 
The effective potential in tortoise coordinates with varying $\alpha$ for the scalar field (left) and EM field (right). Here, $M=1$, $M_{pl}=0.5$ and $l=2$. }
    \label{fig1a}
\end{figure}

\begin{figure}
    \centering
{{\includegraphics[width=8.5cm]{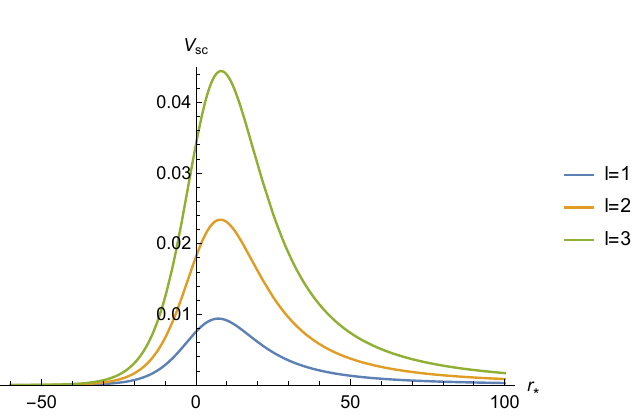} }}\qquad
    {{\includegraphics[width=7cm]{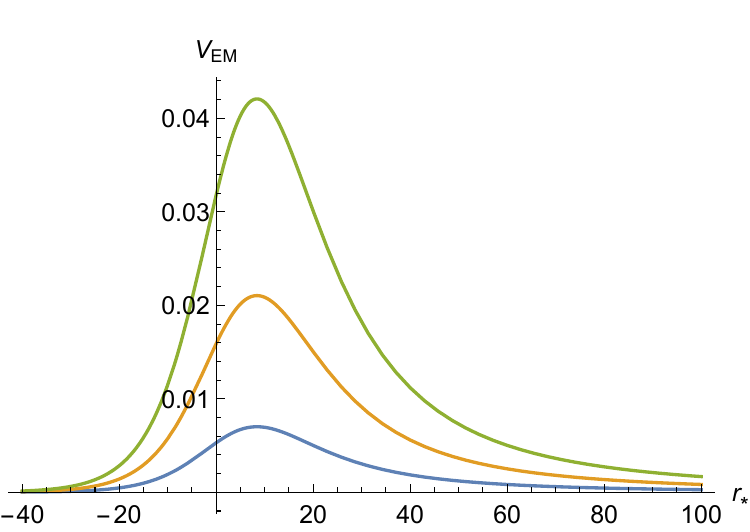}}}
    \caption{ 
The effective potential in terms of tortoise coordinates with varying $L$ for the scalar field (left) and EM field (right). Here, $M=1$, $M_{pl}=0.5$ and $\alpha=0.5$ }
    \label{fig2a}
\end{figure}

In this section, we will consider the linear-quadratic GUP-corrected Schwarzschild black hole to investigate QNMs. The most important step in computing the QNMs of the black hole defined by the metric function (\ref{equ2}) is obtaining the potential associated with the black hole metric. To do so, we must use a probe that is minimally coupled with a scalar field and a Maxwell field described by the equation of motion to perturb the black hole space-time. The general relativistic equations for the scalar $\Phi$ and 
electromagnetic (EM) $A_{\mu}$ fields can be written
as follows:
\begin{equation}
\frac{1}{\sqrt{-g}}\partial _{\mu }\left( \sqrt{-g}g^{\mu \nu }\partial
_{\mu }\Phi \right) =0,  \label{eq21}
\end{equation}
\begin{equation}
\frac{1}{\sqrt{-g}}\partial _{\mu }\left( F_{\alpha \beta }g^{\alpha \nu
}g^{\beta \mu }\sqrt{-g}\right) =0,  \label{eq22}
\end{equation}
where $F_{\alpha \beta }=\partial _{\alpha }A_{\beta }-\partial _{\beta
}A_{\nu }$ is the electromagnetic tensor. Using the separation of the variables to
the above Eqs. (\ref{eq21}) - (\ref{eq22}), then we obtain the Schrödinger wave-like
Form:
\begin{equation}
\frac{d^{2}\Psi}{dr_{\ast }^{2}}+\left( k^{2}-V\right) \Psi=0,  \label{s3}
\end{equation}
where the tortoise coordinate $r_{\ast }$ is defined as $\frac{d}{dr_{\ast }}
=f\frac{d}{dr}$. The effective potential  $V(r)$ for scalar $s = 0$ and EM $s = 1$ fields are
 Given by:
\begin{equation}
V(r
)=f\left(\frac{l\left( l+1\right) }{r^{2}}+f^{\prime }\frac{(1-s^{2})}{r}\right). \label{pot0}
\end{equation}
Plots of the scalar and EM potentials (\ref{pot0}) in terms of tortoise coordinate are produced to investigate the effects of the $l $ and $\alpha$ parameters. Figures \ref{fig1a} and \ref{fig2a} illustrate potential behavior.  In Figure \ref{fig1a}, when $l$ is the same, the potential will increase as $\alpha$ increases, decay, and then disappear, allowing the black hole to return to equilibrium. Similarly, the effective potential increases as $l$ increases when $\alpha$ remains constant. We also observed that the scalar potential is larger than the EM potential. Furthermore, when the tortoise coordinate tends to infinitely, the effective
potentials approach zero.
\\The next step is to determine the QNM frequencies. QNMs are eigenvalues of the wave-like equations (\ref{eq21}) and (\ref{eq22}), corresponding to purely outgoing waves at spatial infinity and purely ingoing waves at the event horizon. We will use the 6th WKB approximation \cite{Konoplya} for efficiency in computing the quasinormal frequencies, which was first presented in Ref. \cite{Iyer}. According to the 6th order WKB approximation, the complex frequency formula takes the form \cite{Konoplya} \begin{equation}
i\frac{\left( \omega ^{2}-V_{0}\right) }{\sqrt{-2V_{0}^{\prime \prime }}}%
-\sum\limits_{i=2}^{6}\Omega _{i}=n+\frac{1}{2}
\end{equation}%
where $V_{0}$  is the maximum of the effective potential $V(r)$, $%
V_{0}^{\prime \prime }=\left. \frac{d^{2}V(r_{\ast })}{dr_{\ast }^{2}}%
\right\vert _{r_{\ast }=r_{0}}$ , $r_{0}$ is the position of the peak value
of the effective potential, and $\Omega _{i}$'s are 2nd to 6th order WKB
corrections that have been given in Refs. \cite{Iyer,Konoplya}. 
The QNMs of the GUP corrected Schwarzschild BH for the massless scalar field perturbation are presented in Table \ref{taba1} with several values of $\alpha$ using the following model parameters:  $M=1$, $M_{pl}=0.5$, and overtone number $n = 0$. Similarly, Table \ref{taba2} presents the QNMs of the EM field perturbation with several values of $\alpha$ using the following model parameters: $M=1$, $M_{pl}=0.5$, and overtone number $n = 0$.
Based on Table \ref{taba1} and \ref{taba2}, it is evident that the linear-quadratic  GUP corrected Schwarzschild BH reaches a stable state after a period of perturbation since the real part of the frequency is positive while the imaginary part is negative. \begin{center}
\begin{tabular}{|c|c|c|c|c|} \hline 
          
$\alpha $ & $l=0$ & $l=1$ & $l=2$ & $l=3$ \\  \hline
$0.25$ & $0.030617-0.033704i$ & $0.0851736-0.028673i$ & $0.141377-0.028323i$
& $0.19755-0.0282373i$ \\ 
$0.50$ & $0.032199-0.035445i$ & $0.0895736-0.030154i$ & $0.148680-0.029786i$
& $0.207756-0.029696i$ \\ 
$0.75$ & $0.033218-0.036567i$ & $0.0924097-0.031109i$ & $0.153388-0.030729i$
& $0.214334-0.030636i$ \\ 
$1.00$ & $0.033912-0.037331i$ & $0.0943403-0.031759i$ & $0.156593-0.031371i$
& $0.218812-0.031276i$ \\ 
$1.25$ & $0.034379-0.037845i$ & $0.0956397-0.032196i$ & $0.158749-0.031803i$
& $0.221826-0.031707i$ \\ 
$1.50$ & $0.034674-0.038170i$ & $0.0964609-0.032472i$ & $0.160112-0.032076i$
& $0.223730-0.031979i$%
\\ \hline 
\end{tabular}
\captionof{table}{The QNM frequencies of massless scalar perturbation with several values of $\alpha$.} \label{taba1}
\end{center}

\begin{figure}
    \centering
    \includegraphics[width=18cm]{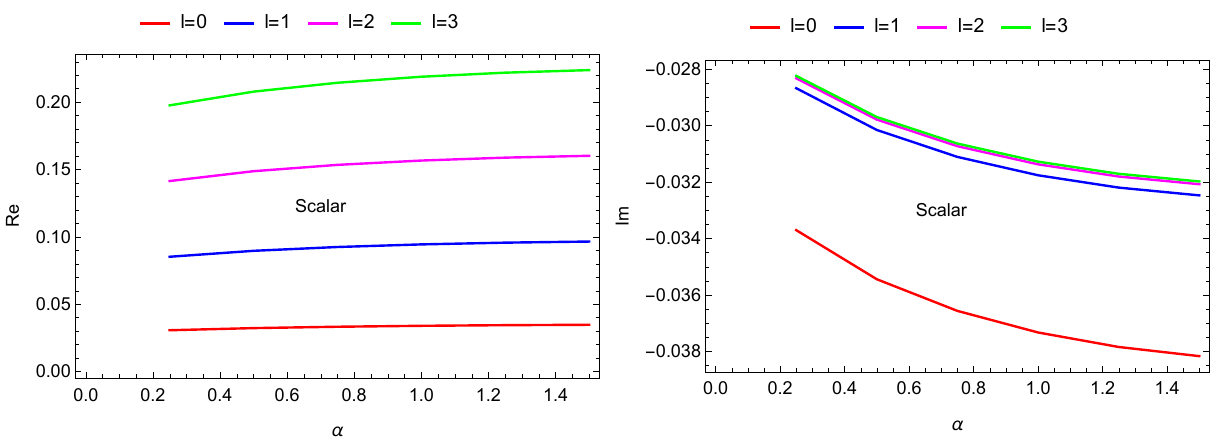}
    \caption{Behavior of the quasinormal frequencies of the scalar field for $l=0,1,2$ and $3$. Here, $n=0$, $M=1$ and $M_{pl}=0.5$.}
    \label{figa3}
\end{figure}

\begin{center}
\begin{tabular}{|c|c|c|c|} \hline 
          
$\alpha $ & $l=1$ & $l=2$ & $l=3$ \\ \hline
$0.25$ & $0.071936-0.027240i$ & $0.133746-0.027814i$ & $0.192146-0.027979i$
\\ 
$0.50$ & $0.075652-0.028648i$ & $0.140656-0.029250i$ & $0.202072-0.029424i$
\\ 
$0.75$ & $0.078047-0.029555i$ & $0.145109-0.030177i$ & $0.208470-0.030356i$
\\ 
$1.00$ & $0.079678-0.030172i$ & $0.148141-0.030807i$ & $0.212825-0.030990i$
\\ 
$1.25$ & $0.080775-0.030588i$ & $0.150181-0.031231i$ & $0.215757-0.031417i$
\\ 
$1.50$ & $0.081469-0.030850i$ & $0.151471-0.031499i$ & $0.217609-0.031687i$%
\\ \hline 
\end{tabular}
\captionof{table}{The QNM frequencies of   EM perturbation with several values of $\alpha$.} \label{taba2}
\end{center}
 To illustrate how the GUP parameter $\alpha$ affects the QNM spectrum, a plot of the real and imaginary QNMs has been made in Figs. \ref{figa3}-\ref{figa4}. When $l$ is the same, Figures show that as $\alpha$ increases, the real and imaginary parts of QNMs increase. This suggests that increasing the  GUP parameter $\alpha$ leads to a longer oscillating frequency and decay rate of the gravitational wave.    On the other hand, when $\alpha$ is the same, as $l$ increases, the real part of QNM significantly increases, and the imaginary part decreases. However, the real part increases slightly, while the imaginary part increases significantly. In general, the results here contrast with those obtained when quantum correction was captured using quadratic GUP \cite{QGUP}. There are striking differences between quantum corrections captured by linear-quadratic GUP and quadratic-GUP. 
 
\begin{figure}
    \centering
    \includegraphics[width=18cm]{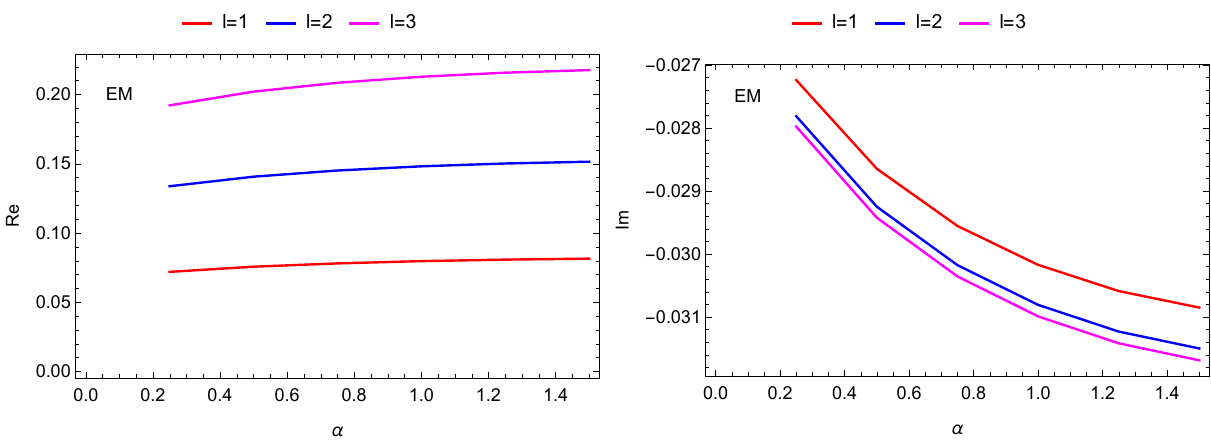}
    \caption{Behavior of the quasinormal frequencies of EM field for $l=1,2$ and $3$. Here, $n=0$, $M=1$ and $M_{pl}=0.5$.}
    \label{figa4}
\end{figure}
\begin{figure}
    \centering
    \includegraphics{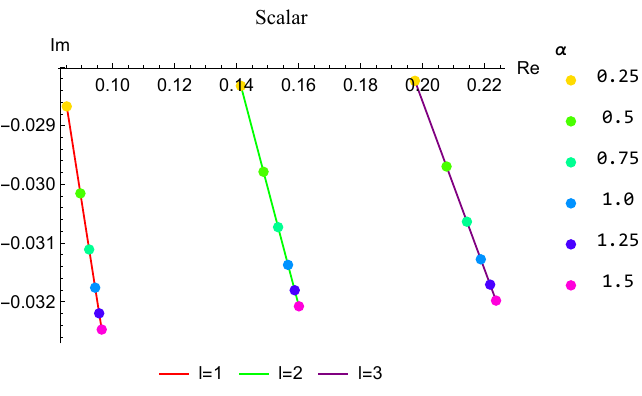}
    \caption{Scalar QNMs in a complex plane for the following modes: $l = 1, 2$ and $3$.}
    \label{figa4}
\end{figure}
\begin{figure}
    \centering
    \includegraphics{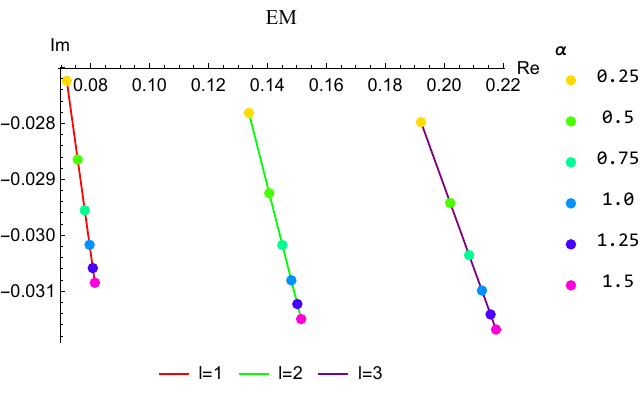}
    \caption{EM QNMs in a complex plane for the following modes: $l = 1, 2$ and $3$.}
    \label{figa4}
\end{figure}
\section{Evolution of scalar and electromagnetic perturbations}
Here, we study the evolution of the scalar and electromagnetic
perturbations using the time domain integration method described
by Gundlach et al.  \cite{gundlach1}.
 It is based on the finite difference method, where the error
 is proportional to $\Delta^2$. Here $\Delta$
  stands for the step used. We can derive the evolution of the field from an
isosceles triangle with the base on the axis r, where the initial
conditions are imposed. Defining $\phi(\tilde{r}, t) = \phi(m
\Delta \tilde{r}, n \Delta t)
 = \phi_{m,n} $, $V(r(\tilde{r})) = V(\tilde{r},t) = V_{m,n}$, we can write:
\begin{equation}
\dfrac{\phi_{m+1,n} - 2\phi_{m,n} + \phi_{m-1,n}}{\Delta
\tilde{r}^2} - \dfrac{\phi_{m,n+1}
 - 2\phi_{m,n} + \phi_{m,n-1}}{\Delta t^2} - V_m\phi_{m,n} + O(\Delta t^2)+ O(\tilde{r}^2)= 0.
\end{equation}
Indexes m and n enumerate, respectively, the coordinates $
\tilde{r}$ and $t$ of the grid:
 $ \tilde{r}_m= \tilde{r}_0 =m\Delta \tilde{r}; t_n=t_0+n\Delta t$.  Now, with a Gaussian
 distribution with finite support as initial condition $\phi(\tilde{r},t)
 = \exp \left[ -\dfrac{(\tilde{r}-\tilde{r}_{md})^2}{2\sigma^2}  \right]$
  and Dirichlet conditions
at $\tilde{r} = 0$,  $\phi(\tilde{r}=0,t)\vert_{t<0} = 0$, time
evolution of the scalar
 field is attempted to compute. Here $\tilde{r}_{md}$ and $\sigma$ are
median and width of the initial wave packet, respectively. Time
evolution of the scalar field can be expressed as
\begin{equation}
\phi_{m,n+1} = -\,\phi_{m, n-1} + \left( \dfrac{\Delta t}{\Delta
\tilde{r}} \right)^2 (\phi_{m+1, n + \phi_{m-1, n}}) + \left(
2-2\left( \dfrac{\Delta t}{\Delta \tilde{r}} \right)^2 - V_m
\Delta t^2 \right) \phi_{m,n}.
\end{equation}
In order to satisfy Von Neumann's stability condition, we need
$\frac{\Delta t}{\Delta \tilde{r}} < 1$ to be kept  maintained.
\begin{figure}[H]
\centering \subfigure[]{ \label{ringingscalar1}
\includegraphics[width=0.3\columnwidth]{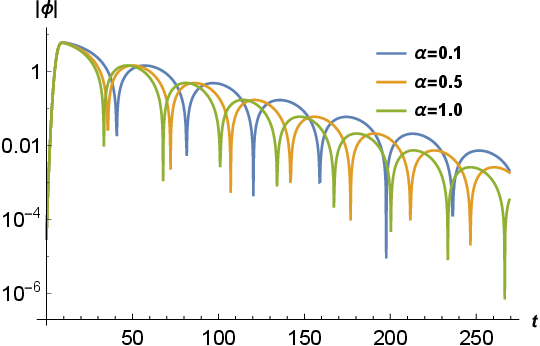}
} \subfigure[]{ \label{ringingscalar2}
\includegraphics[width=0.3\columnwidth]{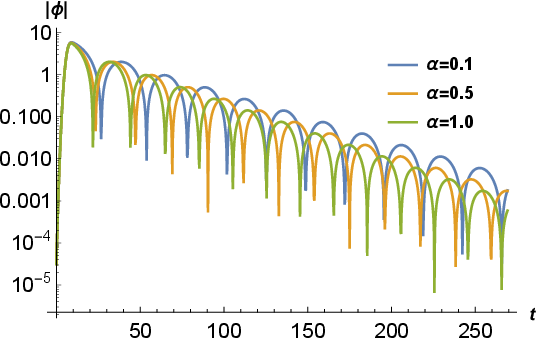}
} \subfigure[]{ \label{ringingscalar3}
\includegraphics[width=0.3\columnwidth]{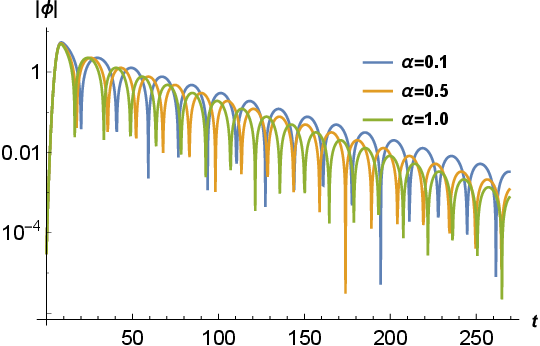}
} \caption{Time domain profile of electromagnetic field for
various values of $\alpha$. The left one is for $\ell=1$, the
middle one is for $\ell=2$ and the right one is for $\ell=3$.}
\label{ringingscalar}
\end{figure}

\begin{figure}[H]
\centering \subfigure[]{ \label{ringingem1}
\includegraphics[width=0.3\columnwidth]{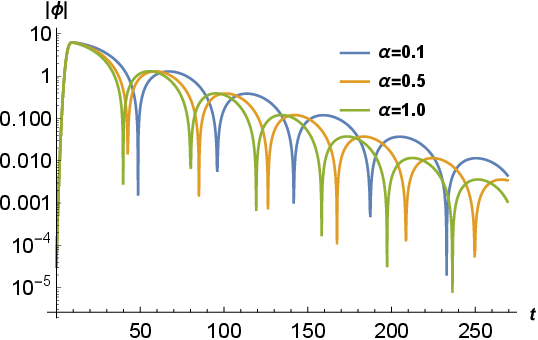}
} \subfigure[]{ \label{ringingem2}
\includegraphics[width=0.3\columnwidth]{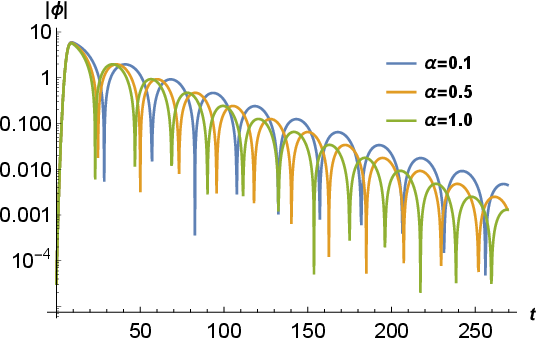}
} \subfigure[]{ \label{ringingem3}
\includegraphics[width=0.3\columnwidth]{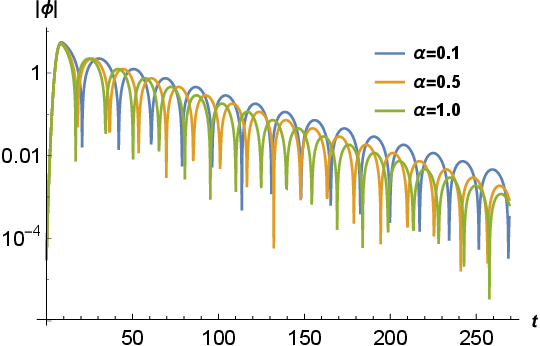}
} \caption{Time domain profile of electromagnetic field for
various values of $\alpha$. Left one is for $\ell=1$, middle one
is for $\ell=2$ and the right one is for $\ell=3$.}
\label{ringingem}
\end{figure}

\begin{figure}[H]
\centering \subfigure[]{ \label{ringing1}
\includegraphics[width=0.4\columnwidth]{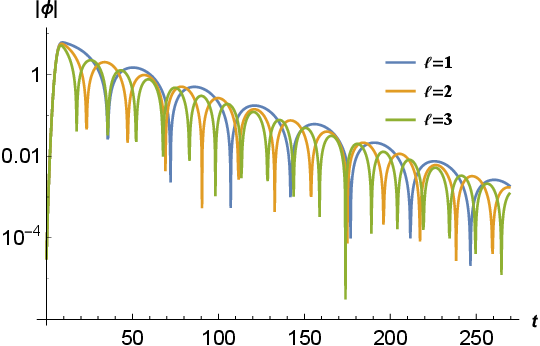}
} \subfigure[]{ \label{ringing2}
\includegraphics[width=0.4\columnwidth]{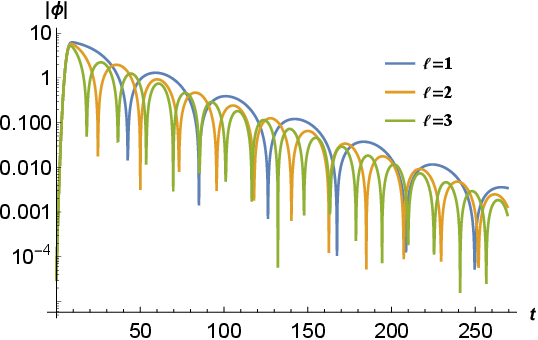}
} \caption{It gives the time profile domain of the field for
various values of $\ell$. The left one is for the scalar field, and the
the right one is for the electromagnetic field. We have taken
$\alpha=0.5$.} \label{ringing}
\end{figure}
The figures above show that the quasinormal ringing for both the
scalar and electromagnetic fields appear to be larger for smaller
value of $\ell$ irrespective of the value of the parameter
$\alpha$. It indicates that the radiation  by the perturbation for
both the scalar and electromagnetic field may be larger for lower
value of $\ell$, and the reverse would be the case for a higher
value of $\ell$.

\section{Greybody Factor}

\begin{figure}
    \centering
{{\includegraphics{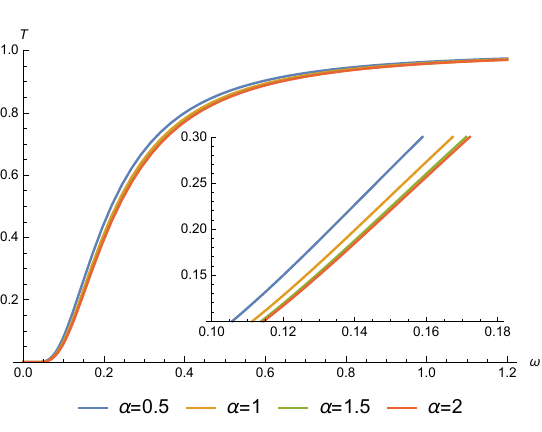} }} \caption{The greybody bound Eq. (\ref{gf1}) of the scalar massless field for several values of the parameters $\alpha$.}
\label{Figa5}
\end{figure}

The goal of this section is to calculate the transmission coefficient or GF for the Schwarzschild black hole with quantum corrections incorporated by the GUP. The GF is a quantity that measures how far the radiation spectrum deviates from black-body radiation. There has already been research conducted on reflection and transmission coefficients (GFs) in multiple scenarios using different methods \cite{qn32,qn36,qn37,qn38,qn39,qn40,qn41,qn42,qn43,qn44,qn45,qn46}. We will use the general semi-analytic bounds method to analyze the GF. This method requires that the GFs be greater than or equal to the following formula \cite{qn41,qn44,qn54} 
\begin{equation}
T\left( w\right) \geq \sec h^{2}\left( \frac{1}{2w}%
\int_{r_{h}}^{+\infty }Vdr_{\ast }\right) .  \label{is8}
\end{equation}
The GF of a massless scalar field will be calculated using the potential given in Eq. (\ref{pot0}).   
 Therefore, Eq. (\ref{is8}) becomes
\begin{equation}
T\left( w\right) \geq \sec h^{2}\left( \frac{1}{2\omega }
\int_{r_{h}}^{\infty }\left( \frac{l\left( l+1\right) }{r^{2}}+\frac{f^{\prime }}{r}\right) dr \right) .\label{in10}
\end{equation}
Analytically, we can solve Eq. (\ref{in10}) as 
\begin{equation}
T\left( w\right) \geq \sec h^{2}\frac{1}{\omega }\left[ \left( -\frac{%
2M^{2}+2l(l+1)M M_{pl}^{2}r_{h}-\sqrt{2}M M_{pl}\sqrt{\alpha }
+M_{pl}^{2}\alpha }{2M M_{pl}^{2}r_{h}^{2}}\right) \right]. 
 \label{gf1}
\end{equation}
Figure \ref{Figa5} depicts the behavior of the obtained GFs by plotting the transmission coefficients versus $\omega$ for various values of $\alpha$. It is obvious that as $\alpha$ increases, the value of the transmission coefficients decreases, implying that less thermal radiation will reach the observer at spatial infinity. 
\section{Spectrum and sparsity of Hawking radiation}
This section elucidates the impact of $\alpha$ on the spectrum and sparsity of the Hawking radiation. The Hawking temperature for the black hole is
\begin{equation}
T_H=\frac{1}{4\pi \sqrt{-g_{tt}g_{rr}}}\frac{dg_{tt}}{dr}|_{r=r_h}=\frac{1}{4 \pi  \alpha  M-8 \sqrt{2} \pi  \sqrt{\alpha } M+32 \pi  M}.
\end{equation}
Variation of the Hawking temperature with respect to $\alpha$ is shown below. It is interesting to observe from Fig. (\ref{hwfig}) that the temperature initially increases with $\alpha$, reaching its maximum value at $\alpha=2.00002$ and then starts decreasing. 
\begin{figure}[H]
\centering
\subfigure[]{
\label{hw1}
\includegraphics[width=0.4\columnwidth]{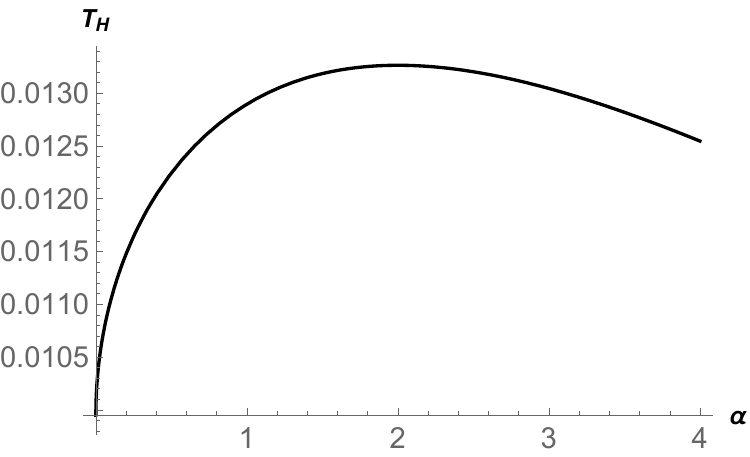}
}
\caption{Variation of Hawking temperature with respect to $\alpha$.}
\label{hwfig}
\end{figure}
Expression for total power emitted by a black hole in the form of Hawking radiation is \cite{yg2017, fg2016}
\begin{equation}
\frac{dE(\omega)}{dt}\equiv P_{tot} = \sum_\ell T (\omega)\frac{\omega}{e^{\omega/T_{H}}-1} \hat{k} \cdot \hat{n}~ \frac{d^3 k ~dA}{(2\pi)^3},
\end{equation}
where $dA$ is the surface element, $\hat{n}$ is unit normal to $dA$, and $T$ is the greybody factor. For massless particles, the above equation yields
\begin{equation}\label{ptot}
P_{tot}=\sum_\ell \int_{0}^{\infty} P_\ell\left(\omega\right) d\omega,
\end{equation}
where $P_{\ell}$, being power spectrum in the $\ell th$ mode, is given by
\begin{equation}\label{pl}
P_\ell\left(\omega\right)=\frac{A}{8\pi^2}T(\omega)\frac{\omega^3}{e^{\omega/T_{H}}-1}.
\end{equation}
$A$ is taken to be the horizon area \cite{yg2017}. To investigate the impact of the GUP parameter on the power spectrum, we plot $P_{\ell}(\omega)$ against $\omega$ for different values of $\alpha$. From Fig. (\ref{pl}), we observe that the peak value of the power spectrum increases with $\alpha$, and the position of the peak value, $\omega_{max}$, shifts towards the right. This observation is true for both perturbations. 
\begin{figure}[H]
\centering
\subfigure[]{
\label{pls}
\includegraphics[width=0.4\columnwidth]{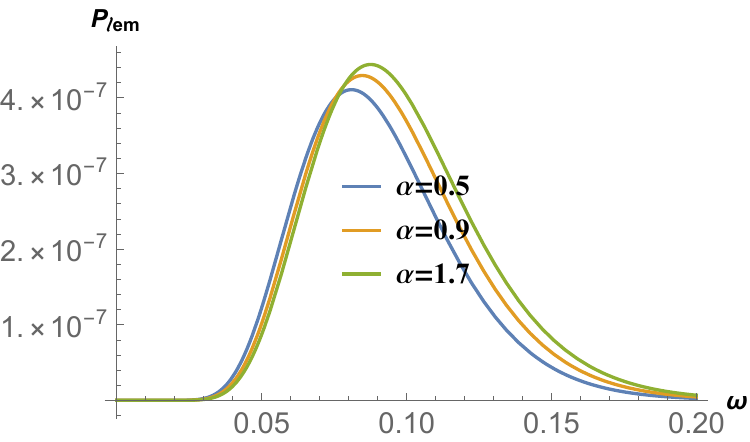}
}
\subfigure[]{
\label{plem}
\includegraphics[width=0.4\columnwidth]{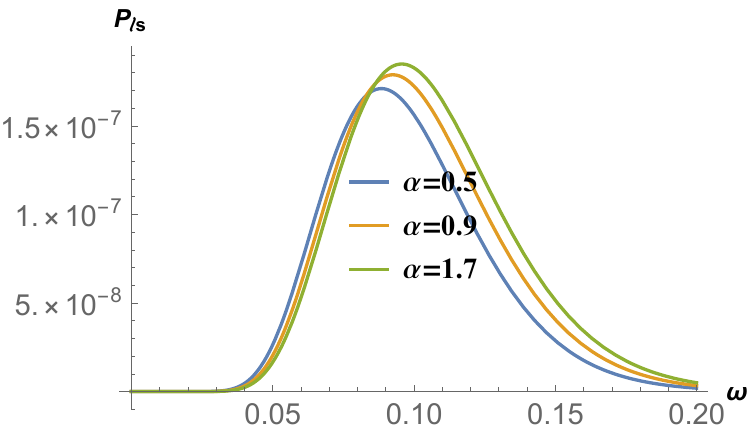}
}
\caption{Power spectrum of the black hole for various values of $\alpha$. The left one is for scalar perturbation, and the right one is for electromagnetic perturbation. Here, we have taken $\ell=2$.}
\label{pl}
\end{figure}
Next, to gauge the sparsity of Hawking radiation, a dimensionless parameter, $\eta$, is introduced and defined by \cite{yg2017, fg2016, ac2020, sh2016, sh2015}
\begin{equation}
\eta=\frac{\tau_{gap}}{\tau_{emission}}.
\label{eta}
\end{equation}
Here, the average time gap between two successive radiation quanta, $\tau_{gap}$,  is given by
\begin{equation}\label{tgap}
\tau_{gap}=\frac{\omega_{max}}{P_{tot}}.
\end{equation}
The expression for the time that is taken by a radiation quantum for emission, $\tau_{emission}$, is given by
\begin{equation}\label{temission}
\tau_{emission} \geq \tau_{localisation}=\frac{2 \pi}{\omega_{max}},
\end{equation}
where $\tau_{localisation}$ is the time period of the emitted wave of frequency $\omega_{max}$. $\eta\ll1$ signifies a continuous flow of Hawking radiation. On the other hand, a large value of $\eta$ implies $\tau_{gap}>>\tau_{emission}$ and a discontinuous flow of radiation. To better understand the impact of $\alpha$ on Hawking radiation, we provide some nuerical values of $\omega_{max}$, $P_{max}$, $P_{tot}$, and $\eta$ for various values of $\alpha$ in Table (\ref{sparsescalar}) for scalar perturbation and in Table (\ref{sparseem}). These values reinforce the conclusions we have drawn from Fig (\ref{pl}). One can also conclude that the total power emitted by the black hole increases with $\alpha$. However, the sparsity fluctuates within a range. Comparing the values of Tables (\ref{sparsescalar}, \ref{sparseem}), we can conclude that the power emitted by the black hole is larger for the electromagnetic perturbation.
\begin{table}
\centering
\caption{Numerical values of $\omega_{max}$, $P_{max}$, $P_{tot}$, and $\eta$ for scalar perturbation for various values of $\alpha$ for $\ell=1$ mode.}
\setlength{\tabcolsep}{-.2mm}
\begin{tabular}{|c|c|c|c|c|c|c|c|c|}
\hline
$\alpha$ & 0.3 & 0.5 & 0.7 & 0.9 & 1.1 & 1.3 & 1.5  \\
\hline
 $\omega_{\max}$ & 0.0854641 & 0.0885899 & 0.0882742 & 0.0929543 & 0.0943996 & 0.0953748 & 0.0953586 \\
\hline
$ P_{\max}$ & $1.64573\times 10^{-7}$ & $1.70930\times 10^{-7}$ & $1.74576\times 10^{-7}$ & $1.78699\times 10^{-7}$ & $1.81100\times 10^{-7}$ & $1.82831\times 10^{-7}$ & $1.84074\times 10^{-7}$  \\
\hline
$ P_{\text{tot}}$ & $1.0404\times 10^{-8}$ & $1.12224\times 10^{-8}$ & $1.18199\times 10^{-8}$ & $1.22684\times 10^{-8}$ & $1.26035\times 10^{-8}$ & $1.28471\times 10^{-8}$ & $1.30145\times 10^{-8}$ \\
\hline
$ \eta$ &  11173. & 111302 & 104924 & 112091 & 112530 & 112689 & 111202  \\
\hline
\end{tabular}
\label{sparsescalar}
\end{table}
\begin{table}
\centering
\caption{Values of $\omega_{max}$, $P_{max}$, $P_{tot}$, and $\eta$ for electromagnetic perturbation for various values of $\alpha$ for $\ell=1$ mode.}
\setlength{\tabcolsep}{-.2mm}
\begin{tabular}{|c|c|c|c|c|c|c|c|c|}
\hline
$\alpha$ & 0.3 & 0.5 & 0.7 & 0.9 & 1.1 & 1.3 & 1.5 \\
\hline
$\omega _{\max }$ & 0.0780561 & 0.0810145 & 0.0833927 & 0.0849507 & 0.0861278 & 0.086868 & 0.0872904 \\
\hline
$P_{\max }$ & $3.95314\times 10^{-7}$ & $4.10567\times 10^{-7}$ & $4.21342\times 10^{-7}$ & $4.29262\times 10^{-7}$ & $4.35082\times 10^{-7}$ & $4.39276\times 10^{-7}$ & $4.42136\times 10^{-7}$ \\
\hline
 $P_{\text{tot}}$ & $2.40713\times 10^{-8}$ & $2.59647\times 10^{-8}$ & $2.73472\times 10^{-8}$ & $2.83849\times 10^{-8}$ & $2.91602\times 10^{-8}$ & $2.97238\times 10^{-8}$ & $3.01111\times 10^{-8}$  \\
\hline
$\eta$ & 40284.2 & 40231.1 & 40472.8 & 40463.8 & 40487. & 40405. & 40274.1 \\
\hline
\end{tabular}
\label{sparseem}
\end{table}
\section{Weak gravitational lensing}
\begin{figure}[H]
\begin{center}
\includegraphics[scale=0.7]{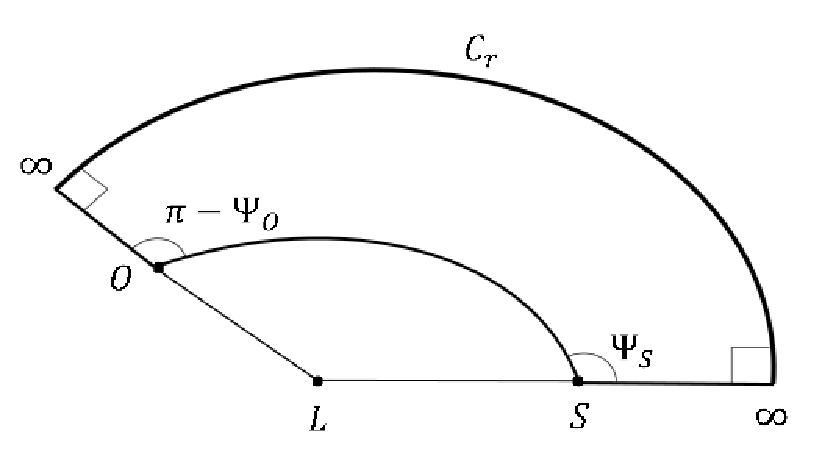}
\caption{Schematic diagram of the quadilateral ${}_O^{\infty}\Box_{S}^{\infty}$. }\label{lensing}
\end{center}
\end{figure}

Gravitational lensing provides an excellent window to probe the effect of the GUP parameter and also differentiates between LG(linear GUP), QG, and LQG on the basis of the astrophysical observable. This section is devoted to the study of the weak gravitational lensing with the LQG-modified black hole acting as a lens (L). We employ the Gauss-Bonnet theorem for our study, which was first proposed by Gibbons and Werner in \cite{GW}. The deflection angle is \cite{ISHIHARA1, CARMO}
\begin{equation}
\gamma_D=\psi_R-\psi_S+\phi_{OS},
\end{equation}
where $\phi_{OS}$ being the angular separation between the observer(O) and the source(S). $\psi_R$ and $\psi_S$ are, respectively, the angles subtended by light rays at the receiver(R) and source. The deflection angle can also be written as
\begin{equation}
\gamma_D=-\int\int_{{}_R^{\infty}\Box_{S}^{\infty}} K
dS,\label{deflectionangle}
\end{equation}
where the quadrilateral ${}_O^{\infty}\Box_{S}^{\infty}$ is shown in Fig. (\ref{lensing}) and K is the Gaussian curvature. For null geodesics, we have 
\begin{equation}
ds^2=0 \rightarrow dt= \pm\sqrt{\zeta_{ij}dx^i dx^j},\label{metric2}
\end{equation}
with
\begin{equation}
\zeta_{ij}dx^i dx^j=\frac{1}{A(r)^2}dr^2+\frac{r^2}{A(r)}\left(d\theta^2+\sin^2\theta\,
d\phi^2\right). \label{metric3}
\end{equation}

The Gaussian curvature, with the help of Eqs. (\ref{metric2}, \ref{metric3}), are given by \cite{WERNER}
\cite{WERNER}
\begin{eqnarray}\nonumber
K&=&\frac{{}^{3}R_{r\phi r\phi}}{\zeta},\\\nonumber
&=&\frac{1}{\sqrt{\zeta}}
\left(\frac{\partial}{\partial \phi}
\left(\frac{\sqrt{\zeta}}{\zeta_{rr}}{}^{(3)}\Gamma^{\phi}_{rr}\right)
- \frac{\partial}{\partial r}\left(\frac{\sqrt{\zeta}}{\zeta_{rr}}{}^{(3)}
\Gamma^{\phi}_{r\phi}\right)\right)\\
&=&\frac{48 M^3-8 M^2 \left(3 \sqrt{2} \sqrt{\alpha }+r\right)+2 M \left(9 \alpha +\sqrt{2} \sqrt{\alpha } r\right)-\alpha  r}{M r^4},\label{gaussian}
\end{eqnarray}
where $\zeta=\det(\zeta_{ij})$. In the above expression, we have put $\hbar=c=1$ and $M_{pl}=0.5$. Now, the deflection angle given in Eq. (\ref{deflectionangle}) can also be expressed as \cite{ONO1}
\begin{equation}
\int\int_{{}_O^{\infty}\Box_{S}^{\infty}} K dS =
\int_{\phi_S}^{\phi_O}\int_{\infty}^{r_0} K \sqrt{\zeta}dr
d\phi,\label{Gaussian}
\end{equation}
where $r_0$ refers to the distance of the closest approach to the
black hole. Assuming straight-line trajectory of the light rays where $r=\frac{b}{sin\phi}$, we first calculate the angle of deflection $\gamma_D^0$ given by 
\begin{equation}
 \gamma_D^0 =\frac{3 \pi  \alpha }{b^2}+\frac{3 \pi  \alpha -8 \sqrt{2} \sqrt{\alpha } b}{2 b^2}+M \left(\frac{64 \alpha }{b^3}+\frac{2 \left(32 \alpha +8 b^2-3 \sqrt{2} \pi  \sqrt{\alpha }
   b\right)}{b^3}\right)+\frac{2 \alpha }{b M}+\mathcal{O}\left( \frac{M^2\sqrt{\alpha}}{b^3},\frac{M^2\alpha}{b^4}\right)
\end{equation}
Then, to obtain higher order corrections, we follow the article \cite{CRISNEJO} where the trajectory is taken to be
\begin{equation}
u=\frac{\sin\phi}{b} + \frac{M(1-\cos\phi)^2}{b^2}-\frac{M^2(60\phi\,
\cos\phi+3\sin3\phi-5\sin\phi)}{16b^3}+\mathcal{O}\left( \frac{M^2\alpha}{b^5}\right)
,\label{uorbit}
\end{equation}
with $u= \frac{1}{r}$, and $b$ is the impact parameter as usual.
Rewriting the integral Eq. (\ref{Gaussian}) as
\begin{equation}
\int\int_{{}_O^{\infty}\Box_{S}^{\infty}} K dS =
\int_{0}^{\pi+\gamma_D^0}\int_{0}^{u}-\frac{K\sqrt{\zeta}}{u^2}du
d\phi.
\end{equation}
we obtain 
\begin{eqnarray}\nonumber
& \gamma_D& \\\nonumber
& =&\frac{16 M}{b}+\frac{24 \pi  M^2}{b^2}+\sqrt{\alpha } \left(\frac{304 \sqrt{2} M^2}{b^3}-\frac{9 \sqrt{2} \pi  M}{b^2}-\frac{4 \sqrt{2}}{b}\right)\\
&&+\alpha  \left(-\frac{4959 \pi  M^2}{16
   b^4}-\frac{392 M}{b^3}+\frac{6 \pi }{b^2}+\frac{2}{b M}\right)+\mathcal{O}\left(\frac{M^3}{b^3}, \frac{M^3\sqrt{\alpha}}{b^4},\frac{M^3\alpha}{b^5}\right)
\end{eqnarray}
Variations of the deflection angle for different types of GUP are shown below. 

\begin{figure}[H]
\centering \subfigure[]{ \label{ringing1}
\includegraphics[width=0.4\columnwidth]{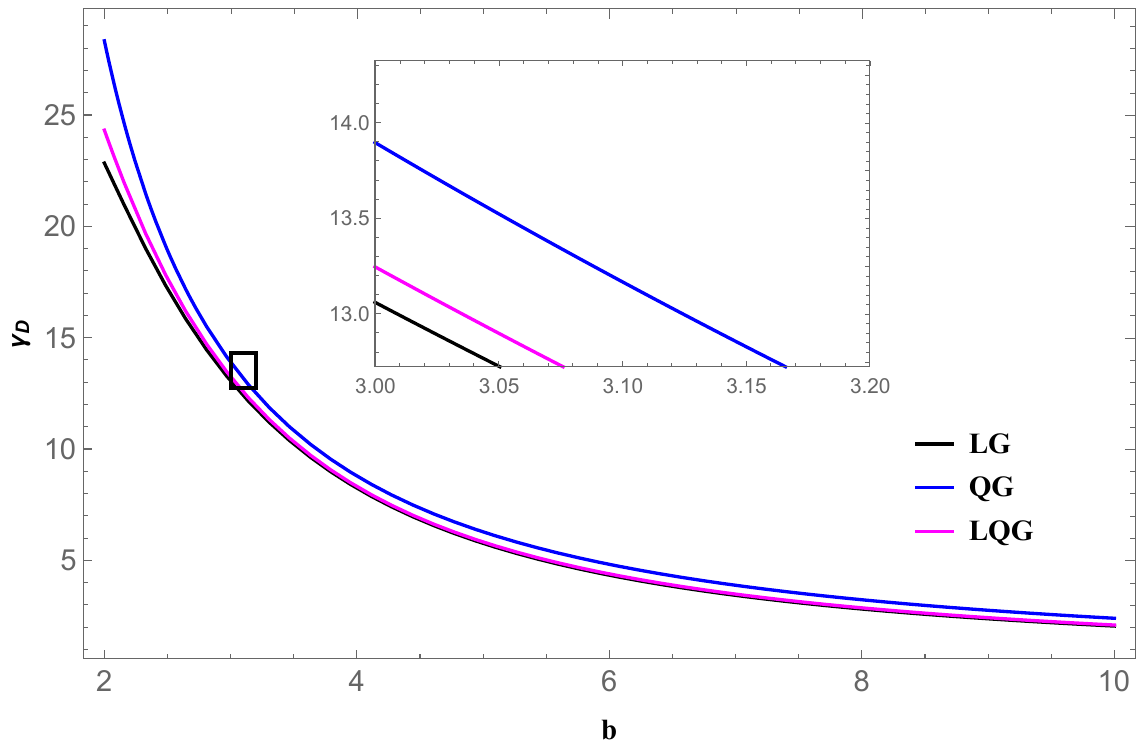}
} \subfigure[]{ \label{ringing2}
\includegraphics[width=0.4\columnwidth]{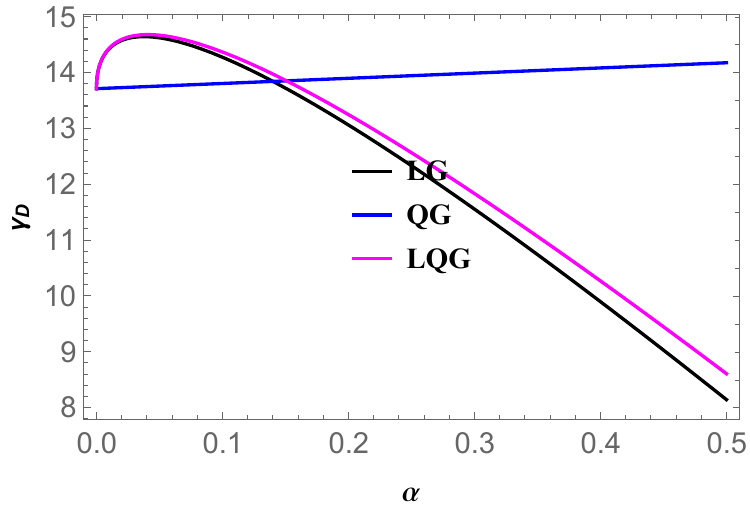}
} \caption{Deflection angle for weak gravitational lensing. The left
panel is for $\alpha=0.2$ and the right panel is for $b=3$.}\label{weak}
\end{figure}
Fig. (\ref{weak}) clearly indicates that the deflection angle is maximum for LQG-modified black holes and least for QG-modified black holes up to a particular value $\alpha$ for any specific value of b, and the reverse is true when we go beyond that particular value of $\alpha$. When we take $b=3$, LQG-modified black holes provide the maximum deflection angle out of three types of GUP-modified black holes for $\alpha \in [0,0.150814)$ and for $\alpha>0.150814$, maximum deflection angle is obtained for QG modified black holes. At $\alpha=0.1508145$, LQG and QG-modified black holes produce the same deflection angle. The nature of variation of the deflection angle with respect to the impact parameter b is the same for all types of GUP. However, it is observed that the deflection angle initially increases with $\alpha$ and then decreases with it for LG and LQG-modified black holes. But the deflection angle increases with $\alpha$ for QG-modified black holes. For LG-modified black holes, the peak occurs at $\alpha=0.0377035$, and for LQG-modified black holes, the peak occurs at $\alpha=0.0407023$ at the impact parameter $b=3$. 
\section{Discussion and Conclusion}
In this endeavor, we have extensively studied LQG-modified static and spherically symmetric black holes. Shadow, QNMs, Hawking radiation, and GL are some astrophysical phenomena that provide excellent avenues to probe the effect of any modification introduced in GR. We have calculated numerical values of shadow radius and constructed shadows with static and infalling accretion. Our study reveals that the shadow size of LQG-modified black holes is larger than that of Schwarzschild black holes. However, LQG-modified black holes have brighter interiors and photon rings than Schwarzschild black holes. \\
Next, we have investigated to study
the quasinormal spectrum for scalar as well as for electromagnetic
perturbations to a LQG-improved Schwarzschild black hole. For this
quantum-improved black hole, we have calculated effective
potential generated through the perturbation of scalar and
electromagnetic field and then computed the frequencies of the
QNMs under scalar and electromagnetic field perturbation using the
WKB method. It is observed that although the variation of the
effective potential with $\ell$ for constant $\alpha$ agrees with
\cite{QGUP}, the variation of it with $\alpha$ keeping $\ell$ to
be fixed shows the opposite picture, and that affects the
quasinormal modes too. Our results show that the real part of the
oscillating frequency increases with the increase in the
LQG-parameter under the scalar field perturbation, whereas the
negative value of the imaginary part of the oscillation frequency
increases with the increase of this parameter. This indicates that
the oscillating frequency and decay rate of the gravitational wave
supposed to emerge under the perturbation seems to be longer with
the increase in the linear-quadratic LQG-parameter $\alpha$. In
\cite{QGUP}, quasinormal ringing has been studied with QG-inspired
quantum corrected black hole, and the results obtained here show
sharp contrast with \cite{QGUP}. A question may naturally be raised: why two
different GUPs are showing contrasting results? We should admit
There is no specific answer to this question right now.
But some comments can be made: the quantum correction has been
here entered in \cite{QGUP} and, in our case, in an indirect manner.
In \cite{QGUP}, it was incorporated through QG. Here, we have used
LQG to capture quantum correction. The real quantum scenario can
be obtained if the quantum theory of gravity is used for the
analysis. Unfortunately, it is not yet developed with its full
wing. Moreover, within the LQG prescription, along with the minimum
measurable length criteria, another important criterion is
involved, which is the maximum measurable momentum that is absent in
the QG prescription.\\
The greybody factor, power spectrum, and sparsity provide insight into the impact of the GUP parameter on thermal radiation as perceived by an asymptotic observer. The greybody factor provides the transmission coefficient of thermal radiation emitted by black holes. Our study shows that the GUP parameter $\alpha$ adversely impacts the transmission coefficient, i.e., less thermal radiation will reach the asymptotic observer. The total power emitted by a LQG modified black hole is observed to be larger than that emitted by a Schwarzschild black hole. The peak value of the power spectrum increases with $\alpha$, and the position of the peak value shifts towards the right. However, the sparsity is found to be fluctuating within a range. These results clearly indicate the significant impact of GUP on thermal radiation.\\
We also delve into studying the deflection angle in the weak field limit. Our aim in this regard is twofold: one is to observe the impact of the GUP parameter on the deflection angle, and the second is to compare the deflection angle for LG, QG, and LQG-modified black holes. Our study reveals a critical value of $\alpha$ for a specific value of impact parameter up to which the maximum deflection angle is observed for LQG-modified black holes. Beyond the critical value, the maximum deflection angle is observed for QG-modified black holes. One example of such a critical value is $\alpha=0.150814$ for the impact parameter $b=3$. At the critical value, the deflection angle for the LQG-modified black hole equals that produced by the QG-modified black hole. Another interesting observation from our study is that the deflection angle for LG and LQG-modified black holes initially increases with $\alpha$ and decreases, but the deflection angle for QG-modified black holes always increases with $\alpha$. The peak of the deflection angle occurs at $\alpha=0.0377035$ for LG-modified black holes and at $\alpha=0.0407023$ for LQG-modified black holes at the impact parameter $b=3$. In this work, we confined ourselves to non-rotaing case. Various aspects of GL and accretion for rotaing case are part of future work. 

\noindent{\bf Acknowledgement:} SKJ acknowledges help from Prof. Ali $\ddot{O}$vg$\ddot{u}$n in the study of accretion. 

\noindent{\bf Data availability Statement:} There is no data
associated with this manuscript. The data are all generated
through numerical computation.

\end{document}